\documentclass[sigconf]{acmart}
\usepackage{multirow}
\renewcommand\footnotetextcopyrightpermission[1]{}
\AtBeginDocument{%
  }

\begin{document}
\title{TempGNN: Temporal Graph Neural Networks for Dynamic Session-Based Recommendations}



\author{Eunkyu Oh}
\email{eunkyu1.oh@samsung.com}
\affiliation{%
  \institution{Samsung Research, Samsung Electronics}
  \city{Seoul}
  \country{Republic of Korea}
}

\author{Taehun Kim}
\email{taehun33.kim@samsung.com}
\affiliation{%
  \institution{Samsung Research, Samsung Electronics}
  \city{Seoul}
  \country{Republic of Korea}
}

\begin{abstract}
Session-based recommendations which predict the next action by understanding a user's interaction behavior with items within a relatively short ongoing session have recently gained increasing popularity. Previous research has focused on capturing the dynamics of sequential dependencies from complicated item transitions in a session by means of recurrent neural networks, self-attention models, and recently, mostly graph neural networks. Despite the plethora of different models relying on the order of items in a session, few approaches have been proposed for dealing better with the temporal implications between interactions. We present Temporal Graph Neural Networks (TempGNN), a generic framework for capturing the structural and temporal dynamics in complex item transitions utilizing temporal embedding operators on nodes and edges on dynamic session graphs, represented as sequences of timed events. Extensive experimental results show the effectiveness and adaptability of the proposed method by plugging it into existing state-of-the-art models. Finally, TempGNN achieved state-of-the-art performance on two real-world e-commerce datasets.
\end{abstract}





\maketitle
\pagestyle{plain}

\section{Introduction}

\begin{figure}[h]
  \centering
  \includegraphics[width=\linewidth]{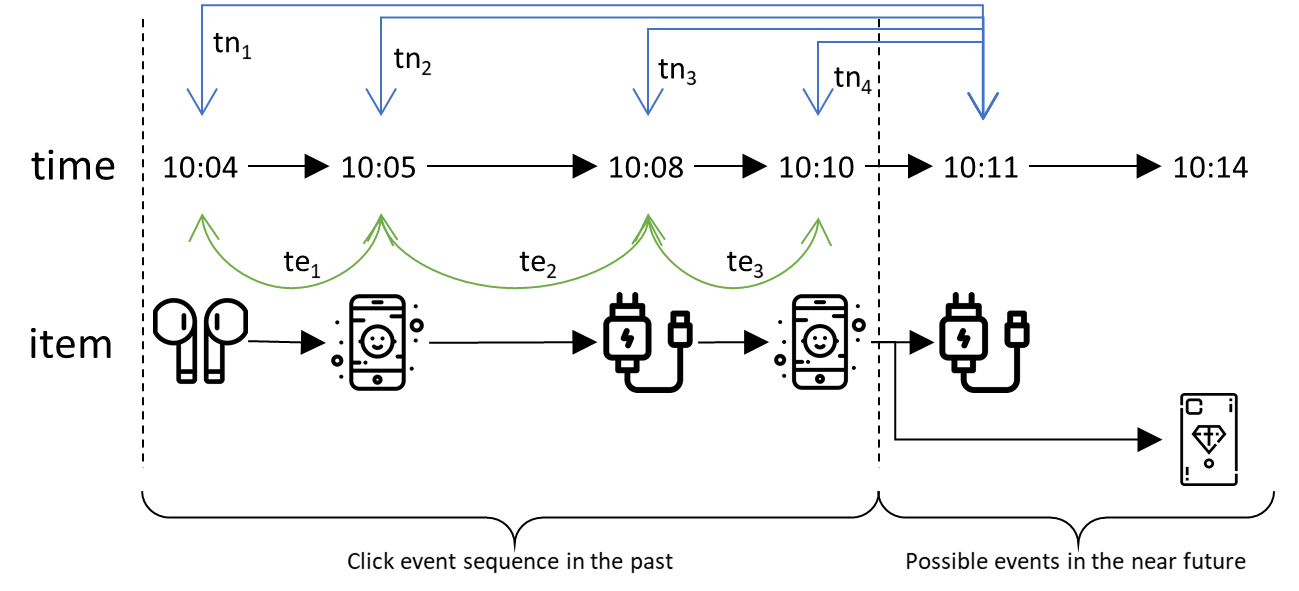}
  \caption{A dynamic session-based recommendation. This aims to address the task of recommending different items according to the timing of the prediction.}
  \label{figure:teaser}
\end{figure}

In the current era of information explosion, recommender systems have been widely adopted in most online e-commerce platforms, news portals, social media, etc. \cite{guy2010social, li2017social, das2017survey, hwangbo2018recommendation} by providing personalized suggestions, so that users can easily find the desired items. However, users' personal information may not always be accessible in some application settings, such as when they browse websites anonymously without logging in. In such cases, only a large number of browsing event sequences from anonymous users can be used. For this, a session-based recommendation (SBR) has gained increasing attention as a method for predicting subsequent items that an anonymous user is likely to interact with given a sequence of previous items consumed in the ongoing session. 

Recent SBR methods have attempted to model how user preferences change over the course of sequential interactions in a session based on deep learning techniques. Recurrent neural network (RNN)-based methods model the series of events in a session as a sequence \cite{hidasi2015session}. In addition, neighbor session information is utilized for augmenting the information on an ongoing session because the information from an ongoing session may be quite sparse \cite{wang2019collaborative}. To determine the effectiveness of long sequences, an attention method has been used for identifying the relevance of each item in a session and capturing the user's main intention \cite{liu2018stamp, li2017neural}. However, blindly feeding the entire series of events into either an RNN or attention-based model makes it difficult to understand the nuanced and intricate structure both within and across sessions, which results in inaccurate modeling. As the most utilized model in recent years, graph neural network (GNN)-based methods convert each session into a directed graph and calculate the degree of information flow between items during information propagation to generate item and session representations \cite{wu2019session, gupta2019niser, pan2020star}. 

Existing methods achieve considerably high performance by identifying pairwise ordered item transition patterns. However, they ignore an important factor, that is, the temporal implications caused by the time difference between events in a session. Most methods assume that all historical interactions have the same importance as the user's current choice, but this may not always be the case. Choices almost always have time-sensitive contexts. The selection of an item by a user is influenced by both short-term and current contexts. Even if it is a click event for the same item, its significance may change depending on when a user clicks on it. Therefore, considering ordered item transitions only without the temporal patterns of the item transition relations is suboptimal for accurately capturing the dynamic changes in user preference. 

Recent efforts have been devoted to considering the temporal implications of events in a sequential recommendation field. \cite{li2020time} models the time intervals between items in a sequence. \cite{zhou2018atrank} bucketizes a time feature with an exponentially increasing time range. \cite{ye2020time} jointly learns user interests and two typical temporal patterns: an absolute time pattern and a relative time pattern. However, such temporal encoding strategies have rarely been considered in SBRs, which restricts their ability to capture the significance of interactions at different times. We conjecture that there are two reasons for the difficulties in inferring temporal meaning from session data. Unlike sequential recommendations, as a session usually has a very short length with a short expiration period, the time differences between interactions in a session are very small. Moreover, because user information is not clearly provided for security reasons, it can be extracted only within the limited time of a session unit.

To tackle the aforementioned problems, we propose Temporal Graph Neural Networks (TempGNN), a generic framework for capturing the structural and temporal dynamics in complex item transitions on dynamic session graphs, represented as sequences of timed events. To facilitate GNNs, we propose a temporal embedding method for nodes and edges. The embedding on nodes models the time differences between the prediction time and timestamps of items, whereas the embedding on edges models the time differences between events in a session. In addition, the proposed temporal encoding approach considers time frequency to avoid biased learning towards a specific time range with high popularity. We also propose a novel method for combining temporal information with an item through a gate network, which allows the model to consider the degree of dependence of time and an item on each other. Extensive experimental results using two real-world e-commerce datasets show that our method outperforms various existing state-of-the-art models and this confirms the effectiveness and adaptability of the proposed temporal embedding.
\begin{figure}[t]
  \centering
  \includegraphics[width=\linewidth]{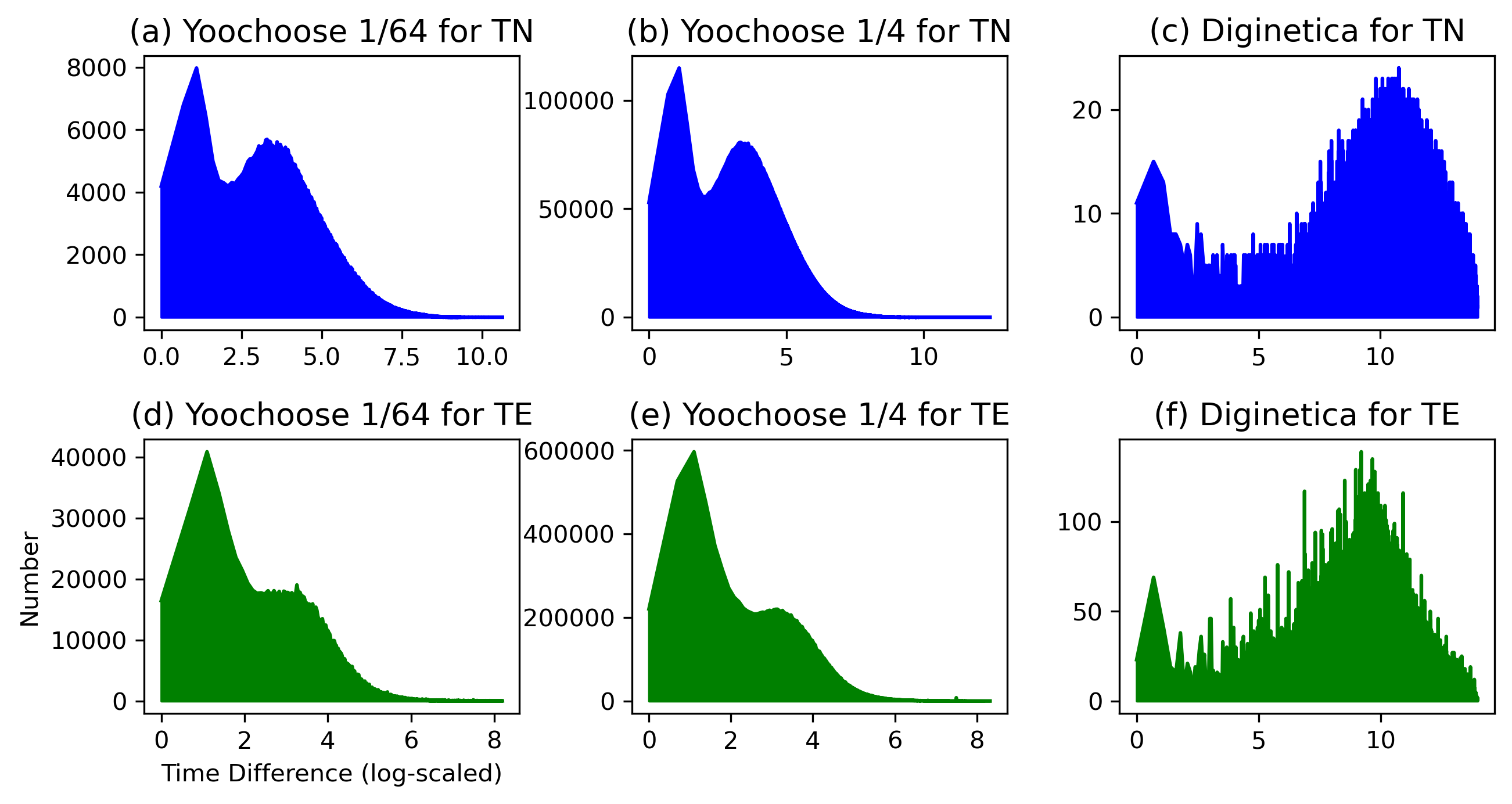}
  \caption{Distributions of time differences in three datasets. The blue distribution indicates time differences from the prediction timing (TN), and the green one is time intervals between interactions in a session (TE).}
  \label{figure:distribution}
\end{figure}

\begin{figure*}[t]
  \centering
  \includegraphics[width=\textwidth]{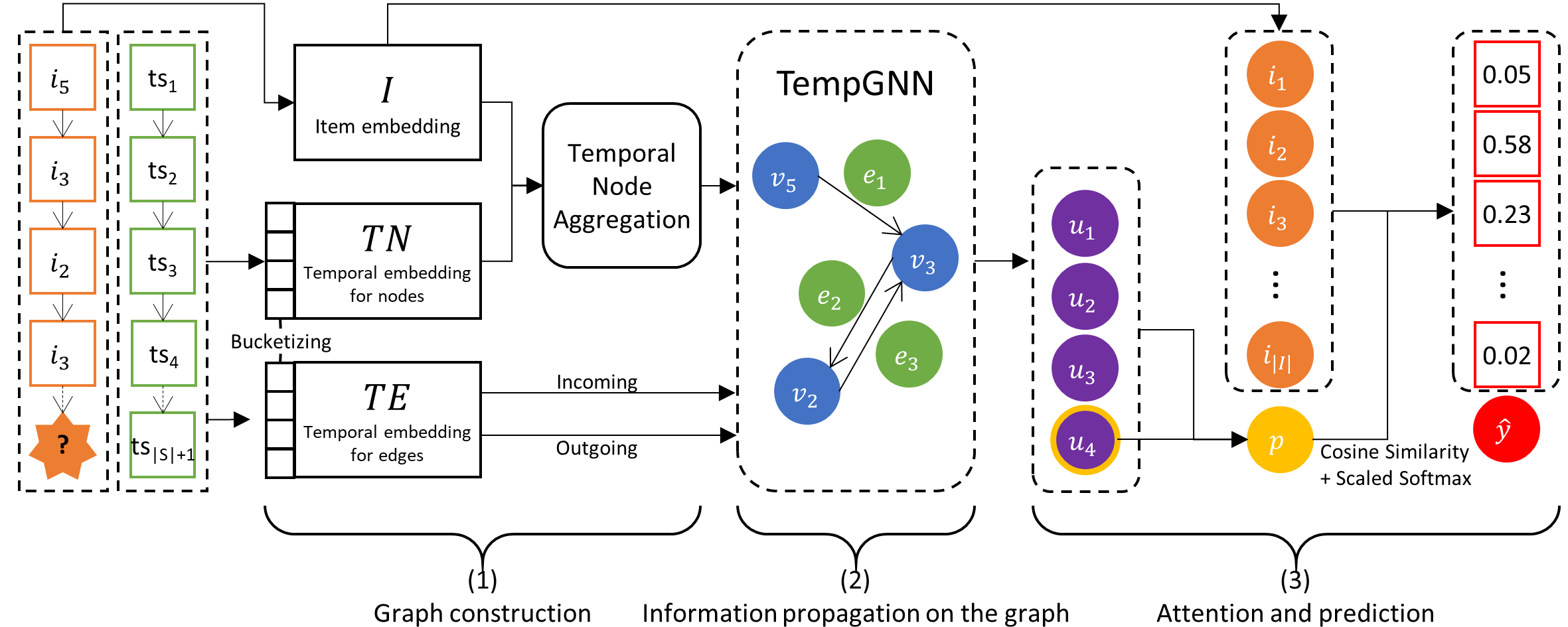}
  \caption{The overall workflow of our model, TempGNN. It consists of three main processes. It takes a prediction timing as input and outputs recommended probabilities for candidate items.}
  \label{figure:ovarll workflow}
\end{figure*}

\section{Related Work}

\textbf{Session-Based Recommendations}. A session is formally represented as a variable-length event sequence with clear boundaries \cite{wang2021survey}. To build better-performing SBRs, a GNN has been widely adopted in recent years for modeling the complex item transitions within and across sessions by transforming a session into a graph structure. SR-GNN \cite{wu2019session} adopted a gated graph neural network (GGNN) \cite{li2015gated} and an attention mechanism for predicting the next item in given a session. NISER+ \cite{gupta2019niser} extended SR-GNN to address popularity bias by introducing L2 normalization. SGNN-HN \cite{pan2020star} introduced a virtual node that considers unconnected items and a highway gate to reduce overfitting. 

However, previous studies still do not perform well in comprehending complicated transition relationships because they do not consider the time differences between the transitions at all. Unlike other methods, our model can capture item transition patterns better by injecting temporal information from a session into GNNs.

\textbf{Temporal Embeddings in Session-based Recommendations}. TA-GNN \cite{guo2020time} utilized time interval information between nodes by introducing time-aware adjacency matrices. This method assumes that two interactions are more relevant when the time interval between them is short. TiRec \cite{zhang2020time} also modeled a session using a time interval. This method adopted a fixed set of sinusoid functions as a basis. KSTT \cite{zhang2021knowledge} introduced three types of time embedding methods that utilize the time difference between each behavior and the prediction time: time bucket embedding, time2vec \cite{kazemi2019time2vec}, and Mercer time embedding \cite{xu2019self}. These multiple temporal embeddings were expected to capture different time patterns.

The above methods attempted to capture a user's intention more accurately using either time intervals between interactions or time differences with regards to the prediction timing, whereas our model adopts both to maximize the effect.

\textbf{Temporal Embeddings in Sequential Recommendations}. Many attempts have been made to encode temporal information in Sequential Recommendations. PosRec \cite{qiu2021exploiting} fully exploited positional information through dual-positional encoding with position-aware GGNNs. 
TGSRec \cite{fan2021continuous} unified sequential patterns and temporal collaborative signals based on Bochner's theorem \cite {loomis2013introduction}. 
However, the small number of learnable vectors of these methods was insufficient to capture a large amount of time information.
Thus, temporal embedding through bucketizing has been widely used. TiSASRec \cite{li2020time} modeled the time intervals between items in a sequence.
ATRank \cite{zhou2018atrank} bucketized a time feature with an exponentially increasing time range.
TASER \cite{ye2020time} jointly learned user interests and two typical temporal patterns in absolute and relative times. However, learning with these methods could be skewed in favor of several buckets with high popularity, because the buckets are divided without considering their frequency of appearance.

In contrast, our time encoding approach additionally considers the time frequency, which ensures that each bucket has the same amount of temporal information to avoid biased learning. This is effective in preventing the overfitting of temporal embeddings, regardless of the distribution of time. In addition, we propose a novel method combining temporal information with an item through a gate network that adjusts each weight by considering the relationship between time and an item.

\section{Problem Definition}
A dynamic session-based recommendation predicts the next item based on an ongoing session taking into account the recommendation timing, as shown in Figure \ref{figure:teaser}. In other words, the next click may change according to the predicted timing. We formulate this task as follows. Let $I= \left\{ i_1, i_2, ..., i_{\left\vert I \right\vert} \right\}$ denote all unique items in all sessions, where $\left\vert I \right\vert$ is the number of unique items. A session $ S = [(i_1^S, ts_1^S), \allowbreak (i_2^S, ts_2^S), \allowbreak ..., \allowbreak (i_{\left\vert S \right\vert}^S, ts_{\left\vert S \right\vert}^S)] $ is a sequence of items and their timestamps, where $i_j^S \in I$ and $ts_j^S$ are the $j$-th clicked item and timestamp in $S$, and $\left\vert S \right\vert$ is the length of the session. Given a session and prediction timestamp $ts_{\left\vert S \right\vert + 1}^S$, we aim to predict the next clicked item $i_{\left\vert S \right\vert + 1}^S$. Items with the highest top-K scores are recommended by estimating a probability vector $\hat{y} \in \mathbb{R}^{\left\vert I \right\vert}$ corresponding to the relevance scores for the unique items.
\section{Method}
Our workflow consists of three main parts: graph construction, information propagation in a GNN, and attention and prediction, as shown in Figure \ref{figure:ovarll workflow}. A detailed description of each process is given in the following sections.

\subsection{Graph Construction}
We construct a graph $G = \left( V, E \right)$ from each session $S$ to better capture user behavior. The vertices of the graph $V = \left\{ v_1, v_2, ..., v_{\left\vert V \right\vert} \right\}$ denote unique nodes from the combinations of items and times.
Edges $E = \left\{ e_1, e_2, ..., e_{\left\vert E \right\vert} \right\}$ are obtained through the temporal embedding of edges.
The dimensions of the embeddings can be set differently, but here they are all set to $d$ for convenience.

\subsubsection{Item Embedding} \label{subsubsection:Item Embedding}
Each unique item is mapped onto a trainable vector $i \in \mathbb{R}^d$ in $I$. We apply $L_2$ normalization to each embedding during training and inference to reduce the effect of the popularity of items on the model because items with a high frequency during training have a high $L_2$ norm, whereas less popular items do not. There are detailed descriptions and experiments in \cite{gupta2019niser}. Therefore, our model uses normalized item embedding as
\begin{equation}
    \tilde{i} = \frac{i}{\lVert i \rVert_2}.
\end{equation}

\subsubsection{Temporal Embedding for Nodes} \label{subsubsection:Time Embedding}
We define temporal embedding for nodes (TN) as a feature of the difference between the prediction timing and click timestamp of each item in a session. This includes information on how much time information should be utilized in predictions, taking into consideration how long ago the interaction was.

Specifically, we first bucketize the time difference. This is a prerequisite for utilizing continuous temporal information, which is difficult to learn. If there are too few buckets, utilizing temporal information has little effect, whereas if the number of buckets is too high, there is a waste of memory with no performance boost. After preparing the appropriate number of buckets, we use a quantile function for the time differences to ensure that each bucket has the same amount of information. This is, of course, performed using the training data. Therefore, the time differences that are not observed during the training process belong to one of the buckets at both ends. After each TN is normalized, it passes through a leaky ReLU function for nonlinearity and a linear layer. In summary, we formulate the $j$-th TN of a session as
\begin{equation}
\begin{aligned}
    & \qquad\; \tilde{tn}_j = W\left( \sigma^{lr} \left( normalize \left(tn_j\right) \right) \right) + b,\\
    & \text{where } tn_j = TN\left[ bucketize^{TN} \left( ts_{\left\vert S \right\vert + 1}^S - ts_j^S \right) \right],
\end{aligned}
\end{equation}
where $W \in \mathbb{R}^{d \times d}$ and $b \in \mathbb{R}^d$ are learnable parameters, $\sigma^{lr}$ is a leaky ReLU function, $normalize$ means $L_2$ normalization, $bucketize^{TN}$ is a function used to obtain a specific bucket index of $TN$, and $TN[]$ is a lookup function that takes one embedding vector corresponding to the index.

\subsubsection{Temporal Node Aggregation} \label{subsubsection:Temporal Node Aggregation}
The nodes used in our GNN reflect the time information to items. Conventional models have used the addition of two embedding vectors as a combination method. The problem with this is that the same temporal information is reflected if the time buckets are the same, regardless of the type of item. However, the relationship between an item and time is more complex. In reality, the degree of sensitivity to time differs depending on the item, even if the temporal embedding is the same. To capture this complex relationship between an item and time, we propose a novel method for controlling the degree of reflection through a gate network when the two embeddings are aggregated, where the weight is calculated by considering the relationship as
\begin{equation}
\begin{aligned}
    &\quad v = \left( 1 - g \right) \odot \tilde{i} + g \odot \tilde{tn},\\
    & \text{where } g = \sigma^{s} \left( W\left[ \tilde{i} ; \tilde{tn} \right] + b \right),
\end{aligned}
\end{equation}
where $\odot$ is an element-wise multiplication, $\sigma^s$ is a sigmoid function, $\left[ ; \right]$ denotes a concatenation, and $W \in \mathbb{R}^{d \times 2d}$ and $b \in \mathbb{R}^d$ are trainable parameters.

\subsubsection{Temporal Embedding for Edges} \label{subsubsection:Edge Embedding}
Our model exchanges information with neighboring nodes by considering their time intervals. This temporal information is very important when exchanging information they have. For example, a time difference between two nodes that is too long might mean they are not adjacent. In addition, a time interval that is too short could mean a miss click within a session. Therefore, we add temporal embedding for edges (TE) to consider temporal information during propagation between adjacent nodes.

We take the timestamp differences of interactions within a session in two directions: incoming and outgoing, and then feature them in the same way as TN in \ref{subsubsection:Time Embedding}. The formula is
\begin{equation}
\begin{aligned}
    & \quad\; \tilde{te}_j = W\left( \sigma^{lr} \left( normalize \left(te_j\right) \right) \right) + b,\\
    & \text{where } te_j = TE\left[ bucketize^{TE} \left( ts_{j + 1}^S - ts_j^S \right) \right],
\end{aligned}
\end{equation}
where $j \in \left\{ 1, 2, ..., \left\vert S \right\vert -1 \right\}$.

\subsection{Information Propagation in a GNN}
Many GNN-based models have been developed for SBRs. Our model advances the previous studies and is particularly based on SGNN-HN and NISER+ \cite{gupta2019niser, pan2020star}. In addition, we utilize temporal information as an additional feature, which can be applied in any recommendation model.

\subsubsection{Star Node}
A star node is a virtual node connected to all nodes in a graph with bidirectional edges. Non-adjacent nodes can also propagate information using the star node as an intermediate node \cite{pan2020star}. At the same time, it has information that integrates the graph. It is updated like other nodes and initialized to the average value of all nodes in the graph as
\begin{equation}
    v_s = \frac{1}{\left\vert V \right\vert} \sum_{j=1}^{\left\vert V \right\vert} v_j,
\end{equation}
where $v_s$ denotes the star node and $\left\vert V \right\vert$ is the number of nodes in the graph.

\subsubsection{Message Passing}
GNNs typically go through step message passing and neighbor aggregation in order to update a node \cite{li2015gated, wu2020comprehensive, niepert2016learning}. Unlike previous GNN-based methods, we designed this exchange of information by considering time intervals between nodes in the message passing phase. Subsequently, a GGNN is applied as a method for updating node information \cite{li2015gated, wu2019session}.

Message passing and aggregation proceed in both directions for incoming and outgoing edges. Among them, we obtain an aggregated message for the $j$-th node from the neighbors through the incoming edges as
\begin{equation} \label{equation:incoming message passing}
\begin{aligned}
    & m_j^I = W^I \left( \frac{1}{\left\vert N_j^I \right\vert } \sum_{v_i \in N_j^I} \left( 1 - g_{ij} \right) \odot v_i + g_{ij} \odot e_{ij} \right) + b^I, \\
    & \qquad \text{where } g_{ij} = \sigma^s \left( W \left[ v_i ; v_j ; e_{ij} \right] + b \right), \\
\end{aligned}
\end{equation}
where $W^I \in \mathbb{R}^{d \times d}$ and $b^I$ are trainable parameters for incoming message passing, $N_j^I$ is the set of incoming neighbors for the $j$-th node, $e_{ij} \in E$ denotes the temporal embedding of the edge from $v_i$ to $v_j$, $W \in \mathbb{R}^{d \times 3d}$ and $b$ are learnable parameters, and the gate $g_{ij}$ considers the characteristics of the two nodes and the time interval between them (i.e., an incoming edge) in order to adjust the transmitted information. The formula for outgoing message passing is similar to Equation \ref{equation:incoming message passing} as
\begin{equation} \label{equation:outgoing message passing}
\begin{aligned}
    & m_j^O = W^O \left( \frac{1}{\left\vert N_j^O \right\vert } \sum_{v_o \in N_j^O} \left( 1 - g_{jo} \right) \odot v_o + g_{jo} \odot e_{jo} \right) + b^O, \\
    & \qquad\qquad \text{where } g_{jo} = \sigma^s \left( W \left[ v_j ; v_o ; e_{jo} \right] + b \right), \\
\end{aligned}
\end{equation}
where $W^O \in \mathbb{R}^{d \times d}$ and $b^O$ are trainable parameters and $N_j^O$ is the set of outgoing neighbors for the $j$-th node. Then, both directional messages are concatenated to update the node as
\begin{equation}
\begin{aligned}
    m_j = \left[ m_j^I ; m_j^O \right].
\end{aligned}
\end{equation}

\subsubsection{Updating a Node}
Updating nodes proceeds by applying the aggregated message vector and star node. First, the message updates the previous information of a node with a gate as
\begin{equation}
\begin{aligned}
    &z_j^l = \sigma^s(W_z m_j^l + U_z v_j^{l-1} + b_z),\\
    &r_j^l = \sigma^s(W_r m_j^l + U_r v_j^{l-1} + b_r),\\
    &\tilde{v}_j^l = \sigma^t(W_h m_j^l + U_h (r_j^l \odot v_j^{l-1}) + b_h),\\
    &\hat{v}_j^l = (1 - z_j^l) \odot v_j^{l-1} + z_j^l \odot \tilde{v}_j^l,
\end{aligned}
\end{equation}
where $W_z, W_r, W_h \in \mathbb{R}^{d \times 2d}$, $U_z, U_r, U_h \in \mathbb{R}^{d \times d}$, and $b_z, b_r, b_h$ are learnable parameters, $l$ denotes the $l$-th layer of the GNN, and $\sigma^t$ is a hyperbolic tangent function. After the propagation of adjacent nodes, the star node is reflected in the update, which considers the overall information in the graph. A gate network helps how much information from the previous star node should be propagated as
\begin{equation}
\begin{aligned}
    & \quad v_j^l = (1 - \alpha_j^l)\hat{v}_j^l + \alpha_j^l v_s^{l-1}, \\
    & \text{where } \alpha_j^l = \sigma^s \left( \frac{\left( \hat{v}_j^l \right)^\top v_s^{l-1}}{\sqrt{d}} \right),
\end{aligned}
\end{equation}
where $v_s^{l-1}$ is the star node of the previous layer in the GNN and $\sqrt{d}$ denotes a scaling factor. A non-parametric mechanism is applied for efficient learning unlike SGNN-HN \cite{pan2020star}. Then, the star node is also updated for continuous graph learning using a non-parametric attention mechanism (i.e., a scaled dot product) as
\begin{equation}
\begin{aligned}
    & \qquad\qquad\quad v_s^l = \left[ v_1^l, v_2^l, ..., v_{\left\vert V \right\vert}^l \right]^\top \beta^l, \\
    & \text{where } \beta^l = softmax \left(\frac{\left[ v_1^l, v_2^l, ..., v_{\left\vert V \right\vert}^l \right] v_s^{l-1}}{\sqrt{d}}\right),
\end{aligned}
\end{equation}
where $\left\vert V \right\vert$ is the number of nodes in a graph and $\left[ v_1^l, v_2^l, ..., v_{\left\vert V \right\vert}^l \right] \in \mathbb{R}^{\left\vert V \right\vert \times d}$ denotes a matrix that includes all nodes in the graph.

\subsubsection{Highway Gate}
This propagation proceeds iteratively with $L$ layers through adjacent and intermediate nodes, which consist of shared parameters. This allows the model to obtain more distant information over multiple propagations. However, the individuality of each node can be diluted if it is excessive. Thus, a highway gate \cite{pan2020star} is applied to take advantage of both as
\begin{equation}
\begin{aligned}
    & \quad v^f = (1 - g) \odot v^L + g \odot v^0, \\
    & \text{where } g = \sigma^s \left( W\left[ v^L ;v^0 \right] + b \right),
\end{aligned}
\end{equation}
where $v^f$ denotes the final node after the highway gate, $v^L$ and $v^0$ are the node after the propagation of the $L$-th layer and initial node, respectively, and $W \in \mathbb{R}^{d \times 2d}$ and $b$ are trainable parameters.

\subsection{Attention and Prediction}
\subsubsection{Obtaining a Preference}
Nodes that have completed all propagations are transformed back to a session format as
\begin{equation}
\begin{aligned}
    U = \left[ u_1, u_2, ..., u_{\left\vert S \right\vert} \right],
\end{aligned}
\end{equation}
where $\left\vert S \right\vert = \left\vert U \right\vert$ is the length of the session and $u_j \in \left\{ v_1^f, v_2^f, ..., v_{\left\vert V \right\vert}^f \right\}$ denotes a node arranged in the original order of the sequence. A representation is obtained by reflecting all nodes in different proportions determined by a soft attention mechanism considering the last and overall information (i.e., a star node) as
\begin{equation}
\begin{aligned}
    &\qquad\qquad\qquad\quad r = \sum_{j=1}^{\left\vert S \right\vert} \gamma_j u_j, \\
    &\text{where } \gamma_j = w_0^\top \sigma^s(W_1 u_j + W_2 u_{\left\vert S \right\vert} + W_3 v_s^L + b),
\end{aligned}
\end{equation}
where $w_0 \in \mathbb{R}^{d}$, $W_1, W_2, W_3 \in \mathbb{R}^{d \times d}$, and $b$ are learnable parameters, $u_{\left\vert S \right\vert}$ is the last node, and $v_s^L$ is the star node after $L$ layers. Because the last node could be a decisive clue for estimating a user's next interaction, a preference vector is formulated as
\begin{equation}
    p = W \left[r ; u_{\left\vert S \right\vert} \right] + b,
\end{equation}
where $W \in \mathbb{R}^{d \times 2d}$ and $b$ are trainable parameters.

\subsubsection{Prediction}
We obtain the normalized probabilities for the next click by measuring the similarities between the preference and all candidate items. To solve the long-tail problem of a recommendation \cite{gupta2019niser}, cosine similarity is applied as
\begin{equation}
    \tilde{y}[j] = \frac{p^\top i_j}{\lVert p \rVert_2 \lVert i_j \rVert_2},
\end{equation}
where $i_j \in I$ is the $j$-th item embedding and $\tilde{y}[j]$, an element of the vector $\tilde{y} \in \mathbb{R}^{\left\vert I \right\vert}$, denotes the similarity between the $j$-th item and a user preference. Then, the similarity vector is normalized by a scaled softmax function, which addresses the convergence problem \cite{gupta2019niser, pan2020star} as
\begin{equation}
    \hat{y}[j] = \frac{\exp \left( \tau \tilde{y}[j] \right)}{\sum_{k=1}^{\left\vert I \right\vert} \exp \left( \tau \tilde{y}[k] \right) },
\end{equation}
where $\tau$ is a scaling factor, $\left\vert I \right\vert$ is the number of candidate items, and $\hat{y}[j]$ denotes the probability that the next click of a user is the $j$-th candidate item. Then, the items with the highest top-K probabilities in $\hat{y} \in \mathbb{R}^{\left\vert I \right\vert}$ are recommended.

\subsubsection{Objective Function}
We adopt a cross-entropy loss function as an objective function for the probabilities. Our model is trained by minimizing the loss, which is formulated as
\begin{equation}
    \mathcal{L} = -\sum_{j=1}^{\left\vert I \right\vert} y[j] log \left( \hat{y}[j] \right),
\end{equation}
where $y[j] \in \left\{ 0, 1 \right\}$ is a target that indicates whether the next click is the $j$-th item or not. In other words, $y \in \mathbb{R}^{\left\vert I \right\vert}$ is a one-hot vector corresponding to the candidate items.
\begin{table}[t]
\caption{Statistics of three datasets.}
\label{table:datasets}
\centering
\resizebox{\linewidth}{!}{
\begin{tabular}{lrrr}
\toprule[1.5pt]
& Yoochoose 1/64  & Yoochoose 1/4  & Diginetica \\
\midrule
\# of clicks & 557,248 & 8,326,407 & 982,961  \\
\# of train sessions & 369,859 & 5,917,745 & 719,470\\
\# of test sessions & 55,898 & 55,898 & 60,858  \\
\# of items & 17,745 & 30,470 & 43,097  \\
Avg. of session lengths  & 6.16 & 5.71 & 5.13  \\
\bottomrule[1.5pt]
\end{tabular}
}
\end{table}

\begin{table*}[t]
\caption{Overall performance for three datasets. A bold-faced number indicates the best score and the second performer is underlined in each column.}
\label{table:overall performance}
\resizebox{\textwidth}{!}{
\begin{tabular}{llcccccccccccc}
\toprule[1.5pt]
\multicolumn{2}{c}{} & \multicolumn{4}{c}{Yoochoose 1/64} & \multicolumn{4}{c}{Yoochoose 1/4} & \multicolumn{4}{c}{Diginetica} \\
\cmidrule(lr){3-6} \cmidrule(lr){7-10} \cmidrule(lr){11-14}
\multicolumn{2}{c}{} & R@20 & M@20 & R@5 & M@5 & R@20 & M@20 & R@5 & M@5 & R@20 & M@20 & R@5 & M@5 \\

\cmidrule{1-2} \cmidrule(lr){3-6} \cmidrule(lr){7-10} \cmidrule(lr){11-14}

\multirow{2}{*}{RNN-based} & GRU4Rec & 62.03 & 23.34 & 37.04 & 20.74 & 67.63 & 27.32 & 42.69 & 24.71 & 34.25 & 9.45 & 14.71 & 7.58 \\
& CSRM & 70.20 & 29.77 & 46.05 & 27.20 & 70.50 & 29.23 & 45.37 & 26.58 & 51.51 & 17.20 & 26.55 & 14.76 \\

\cmidrule{1-2} \cmidrule(lr){3-6} \cmidrule(lr){7-10} \cmidrule(lr){11-14}

\multirow{2}{*}{Attention-based} & STAMP & 68.64 & 29.89 & 45.65 & 27.47 & 70.62 & 30.36 & 46.53 & 27.83 & 47.66 & 15.54 & 24.16 & 13.25 \\
& SR-IEM & 70.86 & 31.59 & 47.95 & 29.16 & 71.02 & 30.49 & 46.69 & 27.92 & 51.70 & 17.14 & 26.46 & 14.66 \\

\cmidrule{1-2} \cmidrule(lr){3-6} \cmidrule(lr){7-10} \cmidrule(lr){11-14}

\multirow{4}{*}{GNN-based} & SR-GNN & 70.38 & 30.71 & 47.08 & 28.26 & 71.39 & 30.96 & 47.07 & 28.40 & 51.46 & 17.54 & 26.94 & 15.11 \\
& NISER+ & 71.36 & 31.91 & 48.21 & \underline{29.46} & 72.74 & 32.09 & \underline{48.82} & 29.55 & 54.39 & 19.20 & 29.15 & 16.70 \\
& SGNN-HN & \underline{71.88} & \underline{31.94} & \underline{48.40} & \underline{29.46} & \underline{72.92} & \underline{32.69} & 48.78 & \underline{30.13} & \underline{55.56} & \underline{19.44} & \underline{29.72} & \underline{16.88} \\
& \textbf{TempGNN} & \textbf{72.60} & \textbf{33.58} & \textbf{48.88} & \textbf{31.09} & \textbf{73.52} & \textbf{34.19} & \textbf{49.62} & \textbf{31.67} & \textbf{56.08} & \textbf{19.96} & \textbf{30.25} & \textbf{17.39} \\
\bottomrule[1.5pt]
\end{tabular}
}
\end{table*}

\section{Experiments}
In this section, we first describe the experimental settings, followed by four experimental results and analyses. All experiments were averaged over five replicates.

\subsection{Experimental Settings}
\subsubsection{Datasets}
\begin{itemize}
\item \textbf{Yoochoose} was released by RecSys Challenge 2015\footnote{\url{https://recsys.acm.org/recsys15/challenge/}}, and contains click streams from an e-commerce website from 6 months.
\item \textbf{Diginetica} was used as a challenge dataset for CIKM Cup 2016\footnote{\url{https://competitions.codalab.org/competitions/11161}}. We only adopt the transaction data.
\end{itemize}

Our preprocessing of these datasets followed previous studies \cite{wu2019session, gupta2019niser, pan2020star} for fairness. We filtered out sessions with only one item and items that occurred fewer than five times. The sessions were split for training and testing, where the last day of Yoochoose and the last week of Diginetica were used for testing. Items that were not included in the training set were excluded from the testing set. Finally, we split the sessions into several sub-sequences. Specifically, for a session $S = [ x_1, \allowbreak x_2, \allowbreak ..., \allowbreak x_{\left\vert S \right\vert}]$, where $x_j = \allowbreak (item, timestamp)$ denotes a pair of items and timestamps, we generated sub-sequences and the corresponding next interaction as $\{[x_1], x_2\}, \allowbreak \{[x_1, x_2], \allowbreak x_3\}, ..., \allowbreak \{[ x_1, x_2, ..., x_{\left\vert S \right\vert - 1}], x_{\left\vert S \right\vert}\}$ for the training and testing sets. As Yoochoose is too large, we only utilized the recent 1/64 and 1/4 fractions of the training set, which are denoted as Yoochoose 1/64 and Yoochoose 1/4, respectively. Statistics for the three datasets are shown in Table \ref{table:datasets}.

\subsubsection{Baselines}
\begin{itemize}
\item \textbf{GRU4Rec} \cite{hidasi2015session} applied gated recurrent units (GRUs) to model sequential information in an SBR.
\item \textbf{CSRM} \cite{wang2019collaborative} employed GRUs to model sequential behavior with an attention mechanism and utilized neighbor sessions as auxiliary information.
\item \textbf{STAMP} \cite{liu2018stamp} applied an attention mechanism to obtain the general preference. 
\item \textbf{SR-IEM} \cite{pan2020rethinking} utilized a modified self-attention mechanism to estimate item importance and recommended the next item based on the global preference and current interest.
\item \textbf{SR-GNN} \cite{wu2019session} adopted GGNNs to obtain item embeddings and recommended by generating a session representation with an attention mechanism.
\item \textbf{NISER+} \cite{gupta2019niser} extended SR-GNN by introducing $L_2$ normalization, positional embedding, and dropout.
\item \textbf{SGNN-HN} \cite{pan2020star} extended SR-GNN by introducing a highway gate to avoid overfitting and a star node, which is a virtual node connected with all nodes. 
\end{itemize}

\subsubsection{Evaluation Metrics}
Following previous studies \cite{wu2019session, gupta2019niser, pan2020star}, we used the same evaluation metrics R@K (recall) and M@K (mean reciprocal rank), where K is 20 and 5. R@K represents the proportion of test instances that have the target items in the top-K recommended items. M@K is the average of the reciprocal ranks of the target items in the recommendation list.

\subsubsection{Parameter Setup}
We used the recent 10 items in a session to ensure fairness across all models. An Adam optimizer was adopted, where the initial learning rate is 0.001 with a decay factor of 0.1 for every 3 epochs, $\beta_1 \!=\! 0.9$, and $\beta_2 \!=\! 0.999$. In addition, the $L_2$ regularization rate was set to $1e^{-5}$. The batch size was 100. All trainable parameters were initialized using a uniform distribution with a range of $\left[ \frac{-1}{\sqrt{d}}, \frac{1}{\sqrt{d}} \right]$ according to the dimension of each model. For our model, the dimension of the embeddings $d$ was set to 256, the scaling factor $\tau$ was 12, and the number of layers $L$ was 6. The numbers of buckets of TN and TE were 40 and 50, respectively. For the other settings of the baselines, we referred to the corresponding paper and official code.

\begin{figure*}[t]
  \centering
  \includegraphics[width=\textwidth]{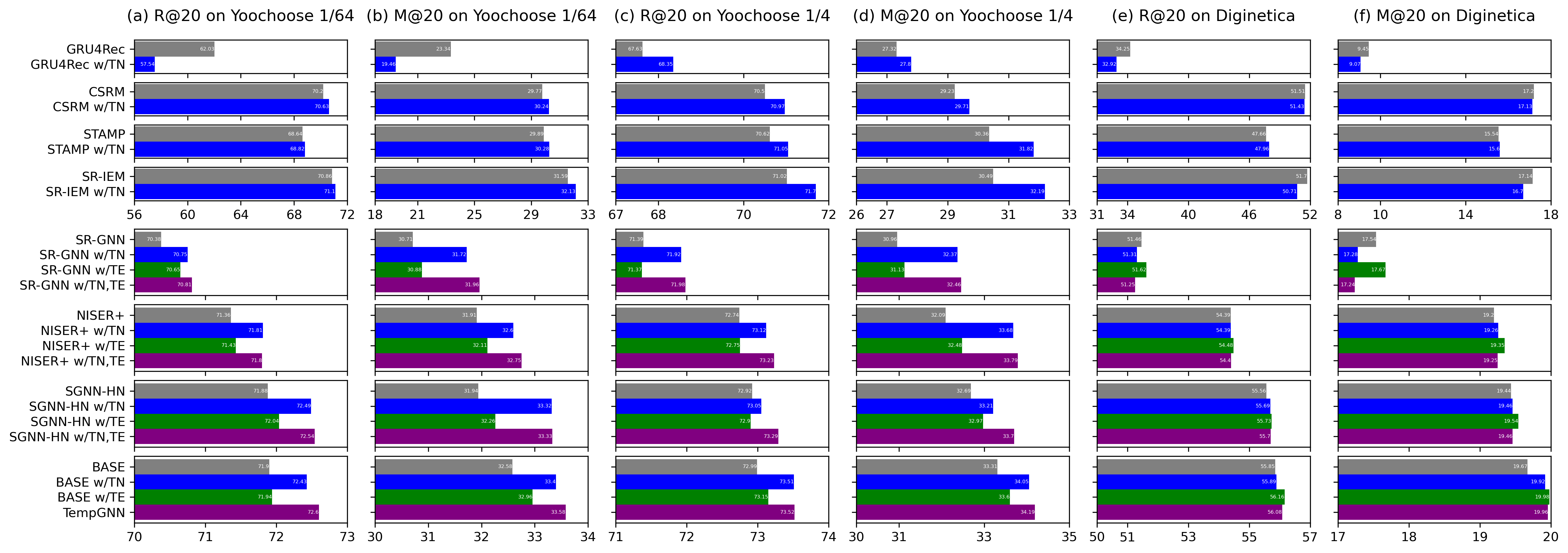}
  \caption{Performance of baseline models with temporal embedding using three datasets. Gray bars indicate the results of basic models. The results of models with TN are shown as blue bars, the ones of models with TE are shown as green bars, and purple bars indicate performance when both are used together.}
  \label{figure:baselines with temporal embedding}
\end{figure*}

\subsection{Overall Performance}
Overall performance comparison of the baselines on the three datasets was summarized in Table \ref{table:overall performance}. This was measured using R@20, M@20, R@5, and M@5. First, if we analyze the difference in the performance of the datasets, it seems that Diginetica is more difficult to predict than Yoochoose. In addition, better performance is obtained when we use a larger training set for Yoochoose. However, the difference of 16 times the learning time and memory usage seems to be not as noticeable.

From among RNN-based models, CSRM outperforms GRU4Rec in all measures due to the addition of an attention mechanism to the GRU. It has better performance than STAMP, except for Yoochoose 1/4, for which both show similar results. Although SR-IEM based on a self-attention mechanism performs similarly to CSRM on Diginetica, it usually outperforms the previous models. SR-GNN, the first GNN-based method to be proposed, has similar performance to SR-IEM overall. NISER+, an extended version of SR-GNN, outperforms all previous results and shows a remarkable performance improvement, especially on Diginetica. SGNN-HN has the second rank performance in most of the measures in this experiment and shows the best performance among the baselines. However, our model outperforms the previous studies in all results. In particular, the improvements in terms of M@20 are notable, which is a measure that is difficult to improve compared to R@20 based on other results. Compared with SGNN-HN, the improvement rates of M@20 on the three datasets are 5.13\%, 4.59\%, and 2.67\%, whereas those of R@20 are 1\%, 0.82\%, and 0.94\%, respectively. Because mean reciprocal rank is a measure that considers the recommendation rank, we can recommend item lists with more sophisticated priorities by adding temporal information.

\subsection{Models with Temporal Embeddings}
The proposed temporal embedding method can easily be adopted in any SBR model. Figure \ref{figure:baselines with temporal embedding} shows the results of utilizing TN, TE, and both together on our model and the baselines described above. Because GRU4Rec, CSRM, STAMP, and SR-IEM among the models are not GNN-based, only TN can be used. The results for the three datasets show different aspects as time information is added.

First, according to the result graphs for Yoochoose 1/64, TN induces an improvement in all models, except for GRU4Rec, which is a pure RNN-based model. This fairly large drop indicates that the model is not trained harmoniously with TN. In addition, although TE does not improve as much as TN, it improves the performance of all GNN-based models. Therefore, the inclusion of temporal information with TN and TE for Yoochoose 1/64 is a very important factor for recommendations.

The results for Yoochoose 1/4 are generally similar to those for Yoochoose 1/64, but even GRU4Rec shows a performance improvement with TN. It appears that a large amount of temporal information leads to improved performance. In addition, TN also shows better results in improvement than TE like the results for Yoochoose 1/64. Interestingly, looking at the results of SR-GNN, NISER+, and SGNN-HN, using only TE does little to improve the performance of R@20, but it helps improve M@20. This means that exploiting the time differences between interactions helps in more sophisticated predictions. In addition, even if TE alone cannot improve R@20, TE with TN yields better results. The results of using the two sets of temporal information together show better performance than those of using TN alone, leading to significant improvements.

The results for Diginetica are different from those for the other two datasets, where the temporal embeddings lead to high improvement rates. In most of the baselines, TN does not help improve, but rather seems to hinder efficient learning. In contrast, TE is helpful for improving all models, which means that the time intervals between interactions in Diginetica provide more clues for accurate predictions. Even the results of GNN-based models show that adding TE alone is better than using both temporal information together.

\begin{table}[t]
\caption{Performance of temporal embedding methods. Q means quantile bucketizing for time, A is an activation function, and G is a gate network when applying temporal embeddings.}
\label{table:temporal embedding}
\resizebox{\linewidth}{!}{
\begin{tabular}{lcccccc}
\toprule[1.5pt]
 & \multicolumn{2}{c}{Yoochoose 1/64} & \multicolumn{2}{c}{Yoochoose 1/4} & \multicolumn{2}{c}{Diginetica} \\
\cmidrule(lr){2-3} \cmidrule(lr){4-5} \cmidrule(lr){6-7}
 & R@20 & M@20 & R@20 & M@20 & R@20 & M@20 \\
\cmidrule{1-1} \cmidrule(lr){2-3} \cmidrule(lr){4-5} \cmidrule(lr){6-7}

Base & 71.90 & 	32.58 & 72.99 & 33.31 & 55.85 & 19.67 \\

\cmidrule{1-1} \cmidrule(lr){2-3} \cmidrule(lr){4-5} \cmidrule(lr){6-7}

Position & 71.86 & 31.84 & 72.88 & 32.24 & 55.68 & 19.43 \\
Constant & 72.27 & 32.57 & 73.38 & 33.75 & 55.85 & 19.76 \\
Bucket & 72.49 & 32.93 & 73.42 & 33.46 & 55.94 & 19.83 \\

\cmidrule{1-1} \cmidrule(lr){2-3} \cmidrule(lr){4-5} \cmidrule(lr){6-7}

Q & \underline{72.57} & 33.44 & \textbf{73.53} & \underline{34.08} & 55.90 & 19.83 \\
Q+A & 72.56 & \underline{33.50} & 73.47 & 34.07 & 56.00 & 19.82 \\
Q+G & 72.43 & 33.43 & \underline{73.52} & \underline{34.08} & \underline{56.07} & \underline{19.91} \\
\textbf{Q+A+G} & \textbf{72.60} & \textbf{33.58} & \underline{73.52} & \textbf{34.19} & \textbf{56.08} & \textbf{19.96} \\

\bottomrule[1.5pt]
\end{tabular}
}
\end{table}

\subsection{Comparison of Temporal Embedding Methods}
Table \ref{table:temporal embedding} shows the results of comparing methods using temporal information for an SBR. Base is a basic version that removes both temporal embeddings (i.e., TN and TE) from TempGNN. Position refers to the model in which positional embedding \cite{vaswani2017attention, qiu2021exploiting} is added to Base. Because only the most recent 10 items are used in our experiment, a maximum of 10 position information can be used. Constant utilizes one learnable vector by multiplying the constant times, which are normalized between 0 and 1. This is similar to the method used in TGSRec \cite{fan2021continuous} except for the periodicity. Bucket means that each bucket is split according to specific intervals after clipping both by 2\% to prevent an outward range. This method has been adopted by many previous models \cite{zhou2018atrank, li2020time, ye2020time}, which do not consider time frequency, whereas our method splits the information into groups of equal size.

The performance after adding positional embedding shows a rather small decrease in the three datasets. The result of Constant shows a slight improvement overall. Although fewer trainable parameters are used compared with Position, the performance is rather improved. This means that clues time differences can provide are more helpful in predicting user behavior from a session than positional differences. The result of Bucket is better than that of Constant, except for M@20 on Yoochoose 1/4, and shows a fairly good improvement compared with Base. Q, which changes the method of allocating buckets from Bucket, shows a high improvement on Yoochoose 1/64 and 1/4, particularly for M@20. Adding an activation function (i.e., a leaky ReLU) results in little change from Q, shown as Q+A, whereas a gate network leads to an improvement on Diginetica, shown as Q+G. Finally, our model's method, Q+A+G, has the best performance, except for R@20 on Yoochoose 1/4. Even so, it exhibits the second-best performance, which is almost equal to the best.

\begin{figure}[t]
  \centering
  \includegraphics[width=\linewidth]{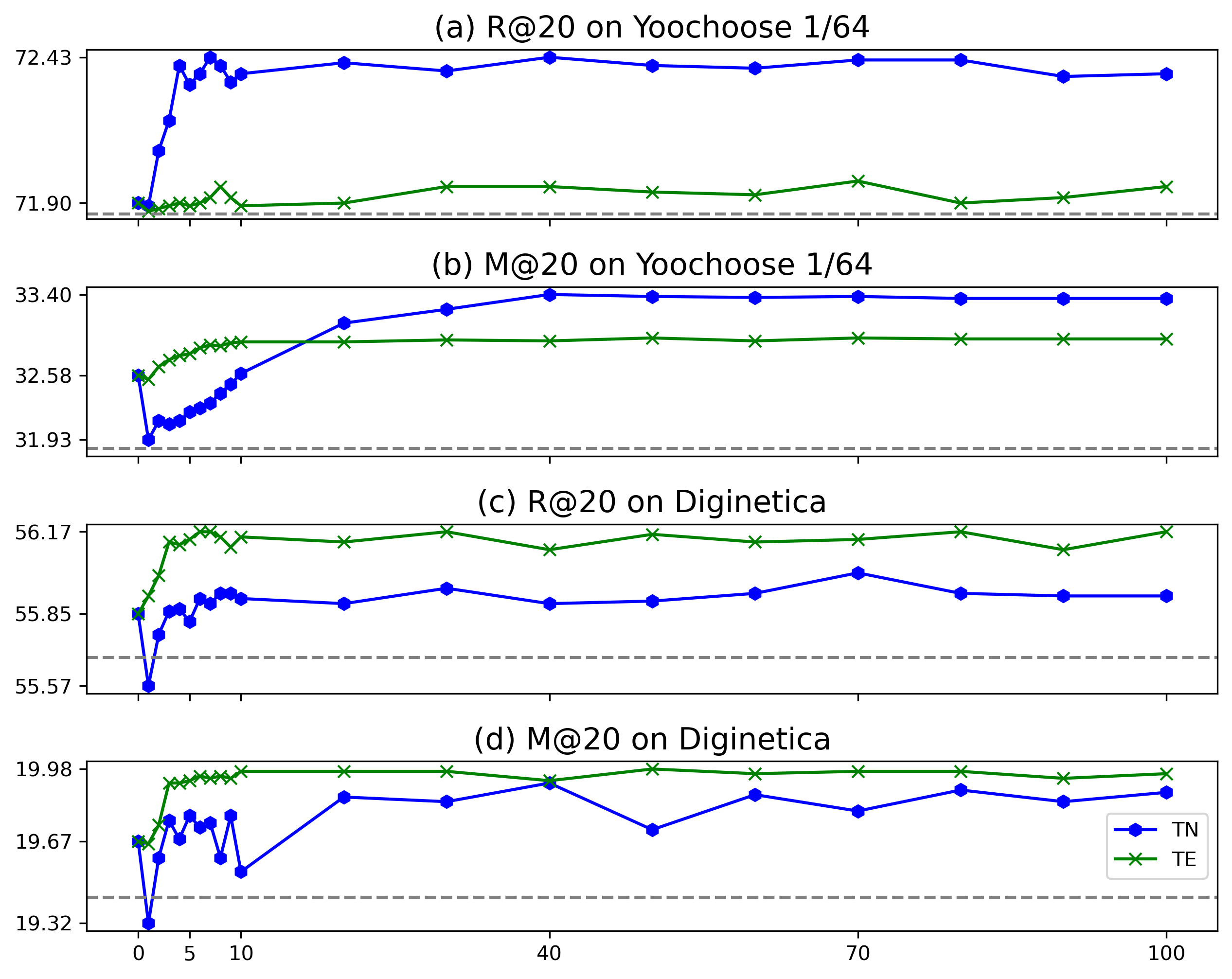}
  \caption{Performance as the number of buckets increases. The gray dotted line indicates the result of positional embedding. Zero buckets indicate a base version without any temporal embedding.}
  \label{figure:the number of buckets}
\end{figure}

\subsection{Number of Buckets}
To maximize the use of time clues in session data, an appropriate number of buckets should be set. This depends on the characteristics of the dataset. If the number of buckets is set too small, the temporal information contained is insufficient. This is because even timestamps with different meanings can be classified into the same group. Conversely, if too many buckets are set, there is a waste of memory due to unnecessary splitting. Even if two timestamps belong to different groups, they would have shown no significant differences. Figure \ref{figure:the number of buckets} shows the number of buckets suitable for TN and TE for the two datasets.

These graphs have three common characteristics. First, too few buckets (e.g., 1 or 2) may degrade the performance compared with a base version without any temporal embedding. This is because unnecessary clues are provided, which actually hinder effective learning. Second, a convergence pattern is exhibited as a certain number is exceeded. This indicates that preparing only a certain number of groups that distinguish times is sufficient. Looking at the converged performance, on Yoochoose 1/64, using TN leads to a meaningful improvement, and utilizing TE does so on Diginetica. Interestingly, both embeddings contribute to improving M@20, which implies that more sophisticated predictions are made. Finally, TN with 10-quantiles always shows higher performance than positional embedding represented by a gray dotted line in the graph. Even if the same number of parameters are used, the difference lies in how the buckets are allocated. Interactions with only one positional gap may be grouped into the same time bucket or there may be a large time difference. Our method shows that it can capture the subtle differences in user behavior that cannot be known by positions.

\begin{figure}[t]
  \centering
  \includegraphics[width=\linewidth]{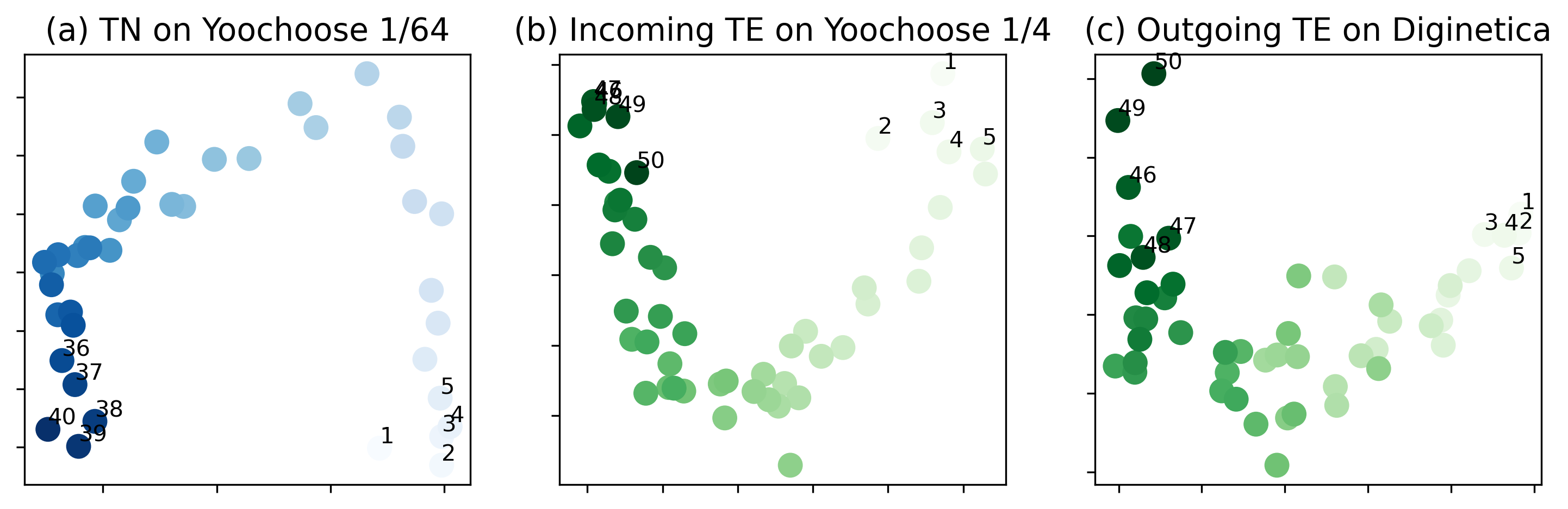}
  \caption{Visualization of temporal embeddings for three datasets. The number of buckets for TN is 40, and 50 for TE. Five indexes at each end of the buckets are annotated.}
  \label{figure:visualization}
\end{figure}

\section{Conclusion}
We introduced TempGNN, a generic framework for capturing the structural and temporal patterns in complex item transitions through temporal embedding operators on nodes and edges on dynamic session graphs represented as sequences of timed events. State-of-the-art results were obtained for several datasets. Extensive experimental results confirm that even if a session has relatively short-length interactions, the temporal relationship between items in the session and the prediction point are important factors in predicting the next item and can improve performance. Meanwhile, although our discrete buckets somewhat reflect continuous distribution over time, as shown in Figure \ref{figure:visualization}, we plan to investigate how to fully capture this in future work.

\bibliographystyle{ACM-Reference-Format}
\bibliography{bibliography}

\end{document}